\documentclass[prb,twocolumn,superscriptaddress,showpacs]{revtex4}
\usepackage[dvips]{graphicx}
\usepackage{latexsym}
\usepackage{bm}
\begin{document}
\newcommand{\fig}[2]{\includegraphics[width=#1]{#2}}
\newcommand{\dprime}{{\prime\prime}}
\newcommand{\be}{\begin{equation}}
\newcommand{\den}{\overline{n}} 
\newcommand{\ee}{\end{equation}}
\newcommand{\bea}{\begin{eqnarray}} 
\newcommand{\eea}{\end{eqnarray}}
\newcommand{\nn}{\nonumber} 
\newcommand{\bk}{{\bf k}}
\newcommand{\vN}{{\bf \nabla}}
\newcommand{\vA}{{\bf A}}
\newcommand{\vE}{{\bf E}}
\newcommand{\vj}{{\bf j}}
\newcommand{\vJ}{{\bf J}}
\newcommand{\bs}{{\bf S}}
\newcommand{\vn}{{\bf v}_n}
\newcommand{\vv}{{\bf v}} 
\newcommand{\la}{\langle}
\newcommand{\ra}{\rangle} 
\newcommand{\ph}{\phi} 
\newcommand{\dg}{\dagger}
\newcommand{\br}{{\bf{r}}} 
\newcommand{\bo}{{\bf{0}}} 
\newcommand{\bR}{{\bf{R}}} 
\newcommand{\bS}{{\bf{S}}} 
\newcommand{\bq}{{\bf{q}}}
\newcommand{\bQ}{{\bf{Q}}}
\newcommand{\vQ}{{\bf{Q}}} 
\newcommand{\hj}{\hat{\alpha}}
\newcommand{\hx}{\hat{\bf x}} 
\newcommand{\hy}{\hat{\bf y}}
\newcommand{\hz}{\hat{\bf z}}
\newcommand{\vS}{{\bf S}} 
\newcommand{\cV}{{\cal U}}
\newcommand{\cD}{{\cal D}} 
\newcommand{\tnh}{{\rm tanh}}
\newcommand{\sh}{{\rm sech}} 
\newcommand{\vR}{{\bf R}}
\newcommand{\crx}{c^\dg(\vr)c(\vr+\hx)}
\newcommand{\crkubox}{c^\dg(\vr)c(\vr+\hat{x})}
\newcommand{\pll}{\parallel} 
\newcommand{\crj}{c^\dg(\vr)c(\vr+\hj)}
\newcommand{\crmj}{c^\dg(\vr)c(\vr - \hj)}
\newcommand{\sumall}{\sum_{\vr}} 
\newcommand{\sumx}{\sum_{r_1}}
\newcommand{\nabj}{\nabla_\alpha \theta(\vr)} 
\newcommand{\nabx}{\nabla_1\theta(\vr)} 
\newcommand{\sumy}{\sum_{r_2,\ldots,r_d}}
\newcommand{\krj}{K(\vr,\vr+\hj)} 
\newcommand{\sigr}{|\psi_0\rangle}
\newcommand{\sigl}{\langle\psi_0 |}
\newcommand{\sier}{|\psi_{\Phi}\rangle}
\newcommand{\siel}{\langle\psi_{\Phi}|}
\newcommand{\sumrj}{\sum_{\vr,\alpha=1\ldots d}}
\newcommand{\krw}{K(\vr,\vr+\hx)} 
\newcommand{\Dtheta}{\Delta\theta}
\newcommand{\rhonew}{\hat{\rho}(\Phi)}
\newcommand{\rhoold}{\hat{\rho_0}(\Phi)} 
\newcommand{\dt}{\delta\tau}
\newcommand{\cP}{{\cal P}} 
\newcommand{\cS}{{\cal S}}
\newcommand{\vm}{{\bf m}} 
\newcommand{\hnr}{\hat{n}({\vr})}
\newcommand{\hnm}{\hat{n}({\vm})} 
\newcommand{\del}{\hat{\delta}}
\newcommand{\upa}{\uparrow} 
\newcommand{\dna}{\downarrow}
\newcommand{\dnk}{\delta n_{\vk}}
\newcommand{\dnks}{\delta n_{\vk,\sigma}}
\newcommand{\dnkp}{\delta n_{\vk '}}

\title{Phase Diagram for Quantum Hall Bilayers at $\nu=1$}

\author{D. N. Sheng}
\affiliation{Department of  Physics and Astronomy, California State
University, Northridge, CA 91330}

\author{Leon Balents}
\affiliation{Department of  Physics, University of California,
Santa Barbara, CA 93106}

\author{Ziqiang Wang}
\affiliation{Department of  Physics, Boston College, Chestnut Hill, MA 02467}

\date{\today}

\begin{abstract}
  We present a phase diagram for a  double quantum well bilayer
  electron gas in the quantum Hall regime at total filling factor $\nu
  =1$, based on exact numerical calculations of the topological Chern
  number matrix and the (inter-layer) superfluid density.  
  We find three phases: a quantized
  Hall state with pseudo-spin superfluidity, a quantized Hall state
  with pseudo-spin ``gauge-glass'' order, and a decoupled composite
  Fermi liquid.  Comparison with experiments provides a consistent
  explanation of the observed quantum Hall plateau, Hall drag plateau
  and vanishing Hall drag resistance, as well as the zero-bias
  conductance peak effect, and suggests some interesting points to
  pursue experimentally.  \typeout{polish abstract}
\end{abstract}


\maketitle

The coexistence of an incompressible integer quantum Hall effect
(IQHE) state and interlayer superfluidity has been established through
a series of experimental and theoretical works on bilayer two
dimensional electron systems at a total electron filling number $\nu
=1$\cite {eisen,wen,murphy,spielman}. Such an IQHE is a consequence of
the strong Coulomb interactions, which lead to a charge gap at $\nu
=1$. Denoting the layer index as pseudospin ``up'' and ``down'', the
ground state is a quantum ferromagnet with spontaneous interlayer
phase coherence\cite{leon1,stern1}, which exhibits superfluidity in the zero
layer separation limit, $d=0$. On the other hand, in the $d\rightarrow
\infty$ limit, the two layers are decoupled, each comprising a
compressible composite fermion liquid at $\nu =1/2$.  The nature of
the phases and transitions\cite{stern2,jog,leon2,numer1,numer2}
between these two limits has attracted many recent studies.  A
generalized pseudospin description suggests a first order
transition\cite{jog}\ at which the pseudospin order vanishes upon
increasing $d$.  The relation of the loss of ferromagnetism with the
disappearance of the IQHE remains unclear, and indeed several
scenarios in which the ferromagnetism and IQHE do not vanish
simultaneously have also been proposed based on the Chern-Simons
mean-field theory\cite{leon2}.  Furthermore,  in real samples
with impurities, a first order transition is believed to be impossible 
on general grounds\cite{imry}. A consistent picture for the phase 
transitions is still absent.

Fundamentally, the distinct phases of the system at $T=0$ are
characterized by their topological order and/or broken symmetries.
Experimentally, these are predominantly reflected in electrical
transport coefficients, particularly the Hall conductance.  Due to the
lack of exact solutions (except at $d=0$ without impurities), the ground
state phases and transport properties have mainly been discussed based
on effective theories\cite{stern2,leon1,leon2}.  Exact numerical
calculations for these systems have been done in the absence of random
disorder potential\cite{numer1,numer2}, which cannot provide direct
information about the transport.  In this letter, we report the first
finite-size exact calculations of transport properties for such a system
in the presence of random impurities, by obtaining the topological Chern
number matrix of the many-body wavefunction and the superfluid density
of the ground state.

The Chern number\cite{chern1,chern2} is a unique integer topological
invariant associated with a wavefunction, and can be used to
distinguish different quantum Hall states.  Physically, the Chern
number equals the boundary condition averaged Hall conductance (in
units of $e^2/h$), so an IQHE state is expected to display a fixed
non-zero integer Chern number independent of disorder configurations.
A state with a non-quantized Hall conductance instead displays a
random integer Chern number intrinsically fluctuating with different
disorder configurations or other external parameters.  Such states are
generally expected to be critical or fluid in nature, since a nonzero
current\cite{chern2,donna} necessarily exists in the bulk to destroy
the exact quantization of Hall conductance.  Thus the distribution of
Chern numbers over samples also reveals the extended or localized
character of the state.

Numerical calculations of the Chern number have up to now only been
carried out from single-particle wavefunctions in non-interacting
systems\cite{chern2,donna}.  The Chern number of a many-body
wavefunction is, nevertheless, well-defined, albeit difficult to
calculate. 
We have developed a new approach to obtain the exact Chern number by
numerically evaluating the Berry phase of the many-body wavefunction
upon changing boundary phases adiabatically\cite{donna1}.  In the
present bilayer system the topological Chern numbers form a $2\times
2$ matrix related to the topological ordering of the system
\cite{topo1}, which determines the charge and spin (we will refer
pseudospin as spin) Hall conductances (as well as the Hall drag, which
is the difference of the two).

\begin{figure}
\begin{center}
\vskip-2.3cm
\hspace*{0.5cm}
\fig{2.2in}{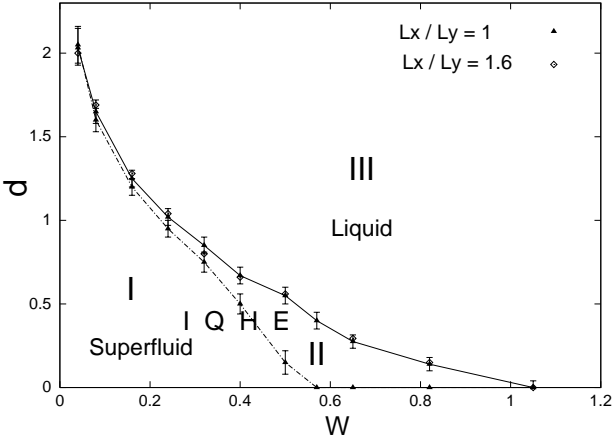}
\vskip-2mm
\caption { The solid line is the critical layer separation $d_c$ vs. $W$, 
which separates the IQHE  plateau  (phase I and II) from the
composite Fermi liquid state (phase III). The dashed line is the 
critical $d_s$ vs. $W$, which separates the superfluid state (phase I) from
the gauge glass state (phase II).}
\label{fig:fig1}
\vskip-8mm
\end{center}
\end{figure}

Our main results can be summarized as a numerical
phase diagram (see Fig. 1) in the $d-W$ plane ($W$ is the disorder
strength) with three distinguishable phases.  Phase I is the usual
bilayer ferromagnet, embodying coexistence of the $\nu =1$ IQHE and
inter-layer superfluidity, occurring at small $d$ and $W$.  At
relatively strong $W$ but small $d$ regime, the superfluid state will
first undergo a phase transition to a ``gauge glass'' (phase II), in
which $n_s$ vanishes due to strong phase frustration. Both phases I
and II have the same Chern number matrix: a uniquely quantized charge
Chern number $C^c=1$ and a random spin Chern number, which result in
nonzero Hall drag conductance.  In phase I, spin superfluidity implies
that the full spin resistivity {\sl tensor} and hence $\rho_{xy}^s$
vanishes below the Kosterlitz-Thouless temperature, so we expect
quantization of the Hall drag resistance,
$\rho_{xy}^d=\rho_{xy}^c-\rho_{xy}^s=h/e^2$ with exponentially small
thermally activated corrections at low temperature. 
Phase II, despite having the same Chern number matrix as phase I,
 has a different behavior of the spin resistivity, since 
the average $n_s$ vanishes. Conventional theory for the gauge glass 
carried over to this problem suggests \cite{leon1}
$\rho_{xy}^s\sim \rho_{xx}^s$ are exponentially small but non-zero at
any $T>0$ and vanish only at $T=0$, leading to a weaker quantization
of the Hall drag resistance.
With the increase of $d$ or $W$, phase II undergoes a transition to
the compressible Fermi liquid (phase III).  Phase III can be
understood from the large $d$ limit of decoupled layers.
It is a metallic state, characterized by zero drag conductance 
and a finite diagonal longitudinal conductance. These results provide 
a consistent understanding for the existing experiments (see below).

In the presence of a strong magnetic field, the Hamiltonian, projected
onto the lowest Landau level, for a double layer two-dimensional
electron gas can be written as: \bea H&=&\sum_{i<j,\alpha ,\beta
}\sum_{{\bf {q}\neq 0}}e^{-q^2/2}V_{\alpha ,\beta
}(q)e^{i{\bf q}\cdot {\bf (R_i^\alpha -R_j^\beta )}} \nn\\
&+&\sum_{i,\alpha }\sum_{{\bf {q}\neq 0}}e^{-q^2/4}V_{{\rm imp}}^\alpha 
(q)e^{i{\bf q%
}\cdot {\bf R_i^\alpha }}
\label{hamil}
\eea
where ${\bf R}_i^\alpha $ is the coordinate of the $i$-th
electron in layer $\alpha $ ($\alpha =1,-1$).
 $V_{\alpha ,\beta }(q)=2\pi e^2/(\epsilon q L_xL_y)
\times
\exp(-qd\delta _{\alpha ,-\beta })$, is the Coulomb potential.
$V_{{\rm  imp}}^\alpha (q)$ is the impurity potential 
generated according to
the correlation relation $<V_{{\rm imp}}^\alpha (q)V_{{\rm imp}%
}^\beta (-q^{\prime })>=\frac {W^2} {L_xL_y}\delta _{\alpha ,\beta }
\delta_{q,q^{\prime }}$,
which corresponds to $<V_{imp}^\alpha ({\bf r})V_{imp}^\beta ({\bf %
r^{\prime }})>=W^2\delta _{\alpha ,\beta }\delta ({\bf {r-r^{\prime }})}$ in
real space, where $W$ is the strength of the disorder.
We set the magnetic length $\ell =1$ and the interaction strength 
$e^2/\epsilon \ell =1$ for convenience. 
We impose the generalized boundary conditions: $ T^\alpha ({\bf L}_j)|\Phi >
=e^{i\theta _j^\alpha }|\Phi >$ ($j=x,y$), to the
finite size double layer system, each in an $L_x\times L_y$ rectangular cell
with an integer number of flux quanta $N_s=L_xL_y/2\pi $. $T^\alpha (
{\bf r})$ is the many-body magnetic translation operator. 
The tunneling term is not considered here in order to
study the interesting limit where the correlation between two layers  is
purely due to Coulomb interaction. We consider up to $N_e=12$ electrons 
at filling number $\nu =N_e/N_s=1$, spanning a Hilbert 
space of size $N_{basis}=853776$.

Through a unitary transformation $\Psi=exp[-i \sum_{i=1}^{N_e}\sum_{\alpha}
(\frac{\theta_x^{\alpha}}{L_x}x_i^{\alpha}+\frac{\theta_y^{\alpha}}
{L_y}y_i^{\alpha})]\Phi$, the topological Chern number\cite{chern1,donna} 
can be calculated as:
\bea
C^{\alpha,\beta}&=&{\frac i{4\pi }}\oint d\theta _j\{\langle {\ \Psi |
{\frac{\partial \Psi }
{\partial \theta _j}}\rangle -\langle {\frac{\partial \Psi }{\partial \theta
_j}}|\Psi }\rangle \}. 
\eea
where $\theta_j$ has layer index $\alpha$ and $\beta$ with
$\theta_j=\theta_{x}^{\alpha}, \theta_y^{\beta}$.
With $\alpha,\beta=-1,1$,  $C^{\alpha,\beta}$ 
forms  a $2\times 2$  matrix.  The closed path integral is 
along the phase boundary of the  $2\pi \times 2\pi $ unit cell.
If common or opposite  boundary phases are opposed on the two layers, then
one obtains charge and spin Chern number, $C^{c,s}$, which is related
to the boundary phase averaged charge and spin Hall conductances $\sigma
_{xy}^{c,s}=C^{c,s}\frac{e^2}h$, respectively. 
We separate the  phase space into a mesh of 64-200
squares. By repeatedly calculating wavefunctions at all nodes 
of the mesh using Lanczos method, we determine the integer
Chern number for the many-body state in each disorder 
configuration.

We first consider the  charge Chern number $C^c$ as a function of
$d$.  At a weak disorder strength $W=0.16$ and $d<1$, 
we find $C^c=1$ for all disorder samples at $N_e=6,8,10,11$ and $12$ 
($20$ samples for $N_e=12$ and $1000$ for $N_e=6$).  
Hence the corresponding ground state displays
the IQHE with total (charge) Hall conductance $\sigma_{xy}^c=\frac{e^2}h$.
As we further increase $d$, a strong fluctuation of the Chern number
takes place at $d\sim 1.1-1.3$, which are caused by level crossings 
upon tuning the boundary
phases and disorder.  The persistence of the crossing of low energy
states at particular values of $d,W$ is a signature of a quantum phase
transition, and is associated with the collapse of the mobility gap.
Thus the mobility of the state in the charge channel is tied
to the fluctuations of $C^c$\cite{chern2,donna}, motivating us to
define $\rho_{ext}=P(C^c\neq 1)$ (the probability of finding $C^c\neq
1$) as a characterization of the extensiveness of
the many-body state.  If $\rho_{ext}$ extrapolates to a nonzero value
in the thermodynamic limit, it represents a fluid phase which can carry
current in the bulk to spoil the exact quantization of $\sigma^c_{xy}$.

\begin{figure}
\begin{center}
\vskip-2.8cm
\hspace*{-3.8cm}
\fig{2.2in}{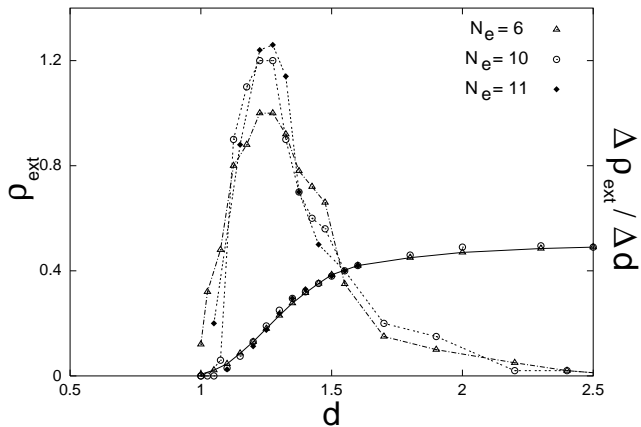}
\vskip-2mm
\caption{ The effective charge mobility $\rho_{ext}$ at $W=0.16$, 
defined as the ratio of the number of samples with $C^c \neq 1$ and the 
total number of samples, as a function of $d$ at $N_e=6$, $10$ and $11 $. 
$\Delta \rho_{ext}/ \Delta d$ vs.  $d$ is also shown. }
\label{fig:fig2}
\vskip-8mm
\end{center}
\end{figure}

To  determine the critical  $d_c$ for
the $\nu =1$ IQHE to metal transition, we plot $\rho_{ext}$ as a
function of $d$ at $W=0.16$, $N_e=6$, $10$ and $11$ in Fig.2. $\rho
_{ext}$ always increases rapidly around $d\sim 1.1-1.3$ and saturates
to 0.5  at larger $d$.  In the same figure, we plot $\Delta
\rho_{ext}/\Delta d$ ($\Delta d=0.05$) as a function of $d$. Around
$d=1.25$, there is always a very strong peak growing with $N_e$.  We
determine the critical layer separation $d_c$ ($=1.25\pm0.05$ for
$W=0.16$) for the IQHE to metal transition by the location of the
maximum of this peak for each $W$.  A step jump of $\rho_{ext}$ is
expected for such a transition in the thermodynamic limit.

The critical $d_c$ as a function of $W$ determined in this manner is
indicated in Fig.~1 by the solid line. The $d_c$ for two aspect ratios
($L_x/L_y=1$ and $1.6$) agree well with each other, demonstrating the
smallness of overall finite size effects. At $W\rightarrow 0$, $d_c$
saturates to around $d_c=2.2\pm 0.2$ with relatively large error bars due to
increasing finite-size effects at very weak $W$ (at $W=0.04$,
$d_c=2.02\pm 0.13$).  We note that studies based on the local ferromagnetic
moment $m_{FM}$ for weak disorder will tend to underestimate $d_c$, due to a
large reduction of $m_{FM}$ by low energy states mixing around $d\approx
1.5$, which, however, does not affect the Chern number.

More information on the nature of spin sector can be obtained by
tuning the boundary conditions in two layers according to
$\theta _t=\theta _x^1=-\theta _x^{-1}$.
The energies of the lowest two states $E_g$ and $E_1$ 
change significantly with $\theta_t$, with variations of the order 
of the level spacing.  At small $d$ and $W$, $E_g$ first 
increases as a quadratic function of $\theta_t$ until it is 
energetically favorable for a vortex (through the hole in the torus 
encircled by the $x$ axis) to enter the system, a typical feature of 
a superfluid state.  
The disorder averaged superfluid density  can be calculated as $n_s=
\frac 1 2 \langle \left.\frac{\partial ^2
    E_g}{\partial\theta_t^2}\right|_{\theta_t=0}\rangle$\cite{rhos}.  

\begin{figure}
\begin{center}
\vskip-1.8cm
\hspace*{-1cm}
\fig{2.2in}{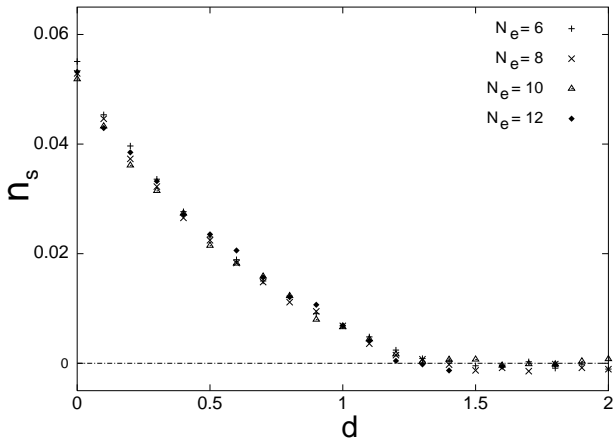}
\vskip-5mm
\caption { The superfluid density $n _s$ vs. $d$ at $W=0.16$. At the
transition point $d_s=1.2$, $n _s \rightarrow 0$. }
\label{fig:fig3}
\vskip-1.cm
\end{center}
\end{figure}

In Fig. 3, $n_s$ is plotted as a function of $d$ at $W=0.16$ for $
N_e=6,8,10$ and $12$. The overall behavior agrees with the generalized
mean-field calculation for pure system (our definition of $n_s$ is twice
that of Ref.~\onlinecite{jog}). Naturally, $n_s$ reduces with increasing 
$d$, and we define the boundary of the superfluid 
state $d_s$ by $n_s=0$, e.g. $d_s=1.2$ for $W=0.16$ shown here. The 
critical $d_s$ is also shown in Fig. 1 as the dashed line. In the strong 
$W$ case, $d_s$ becomes obviously smaller than $d_c$, indicating a superfluid
state (phase I) to phase II transition inside the IQHE regime.
In phase II, the {\sl averaged} superfluid density vanishes, due
principally to phase frustration: $E_g(\theta_t)$ still depends 
strongly upon $\theta_t$ with positive or negative curvatures 
depending on disorder realizations.  We postulate hence
that the spin sector in phase II behaves as a gauge or vortex
glass\cite{matthew}, with Edwards-Anderson magnetic order and algebraic
stiffness $\langle |E_g(\theta_t=\pi/2)-E_g(\theta_t=0)|\rangle \sim
L^{-|\Theta|}$, as proposed for $\nu=1$ bilayers in
Ref.~\onlinecite{leon1}.

Further evidence that phase II is not a spin insulator is obtained
from the spin Chern number $C^s$, determined by imposing opposite
boundary phases  to both layers.  We find that $C^s$
fluctuates (around 1) throughout the $d-W$ plane, implying fluidity 
of the spin sector and a non-quantized spin
Hall conductance.  This rules out a spin insulator for phase II,
favoring the interpretation as a gauge glass. 
In both phase I and II, the Hall drag
conductance $\sigma_{xy}^{1,-1}+\sigma_{xy}^{-1,1}=\frac{1}{2}
(\sigma_{xy}^{c}-\sigma_{xy}^s)
=(C^c-C^s)\frac{e^2}{2h}$, is also 
non-quantized and non-zero, which is a consequence of the coupling
between two layers.  At the phase boundary for the IQHE ($d=d_c$), we find
that the nondiagonal Chern number $C^{1,-1}=C^{-1,1}=0$ and the drag
Hall conductance drops to zero, indicating that the spin sector is
also involved in the phase transition at $d=d_c$.  At this gauge glass
to composite Fermi liquid critical point, we expect the spin
correlations go from (Edwards-Anderson-)superfluid to metallic.

\begin{figure}
\begin{center}
\vskip-1.8cm
\hspace*{-1cm}
\fig{2.2in}{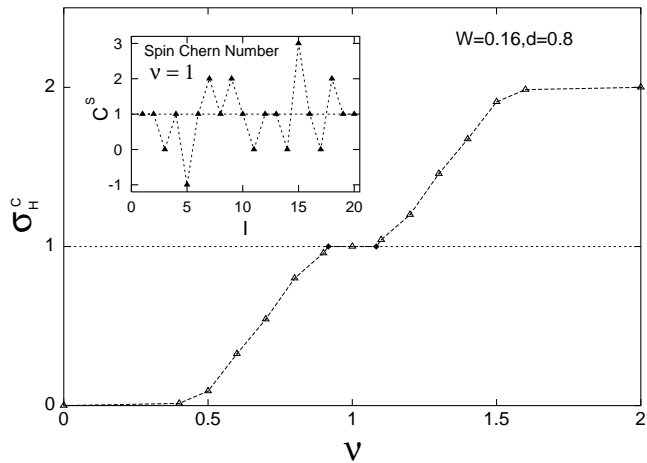}
\vskip-0.5cm
\caption{ The charge Hall conductance $\sigma_{xy}^c$ (in units of $e^2/h$) 
vs.  filling number $\nu =N_e/N_s$.
In the inset, the random spin Chern number $C^s$ 
vs. disorder configuration $I$ for 20 samples at $\nu=1$.
$W=0.16$ and $d=0.8$ (in phase I) for both data.}
\label{fig:fig4}
\vskip-8mm
\end{center}
\end{figure}

To reveal the charge plateau, we calculate $\sigma_{xy}^c$ as a
function of $\nu$.  Shown in Fig.4, at $W=0.16$ and $d=0.8$, we
observe an exact quantized plateau between $\nu =11/12=0.91$ and $\nu
=13/12=1.09$, with $\rho_{ext}=0$ corresponding to
a finite mobility gap.  The plateau width is usually smaller or around
$\Delta \nu =0.2$ depending on $W$, in good agreement with 
experiments\cite{experi}.  In contrast to the plateau in charge channel 
around $\nu=1$, the
spin Chern number  fluctuates with different
disorder configurations, as shown in the inset of Fig.  4.

We conclude with some comparison to experiments.  Both phases I and II
exhibit the IQHE in the charge channel, i.e. $\rho_{xy}^c=h/e^2$ and
$\rho_{xx}^c=0$ at zero temperature, and we expect activated
corrections at $T>0$.  The spin Chern number is random and
fluctuating, which gives rise to a nonquantized spin conductance.
The exact quantized Hall plateau (for the charge) has been observed by
several experimental groups\cite{eisen,experi} with a critical
$d_c^{exp}=1.8$, corresponding to the regime of very weak $W\approx
0.07$ in our phase diagram.  In this regime, the 
width of phase II is extremely narrow if non-zero, and the Hall
plateau phase at $d<d_c$ observed experimentally therefore corresponds
to phase I.  As a true two-dimensional superfluid, it is expected to
exhibit anomalous properties, e.g. divergent $\sigma_{xx}^s$ {\sl even
  at non-zero temperatures below the Kosterlitz-Thouless temperature,
  $T<T_{KT}$}.  The experimentally observed zero-bias tunneling
conductance peak below a characteristic temperature 
\cite{spielman}\ is a direct reflection of the
associated off-diagonal order in the spin channel.  Furthermore, the
divergent spin conductivity implies that the full spin resistivity
tensor must vanish, $\rho_{xx}^s=\rho_{xy}^s=0$ for $T<T_{KT}$.  At
$T=0$, the IQHE then implies quantized Hall drag
$\rho_{xy}^d=\rho_{xy}^c-\rho_{xy}^s=h/e^2$ and vanishing longitudinal
drag, $\rho_{xx}^d=\rho_{xx}^c-\rho_{xx}^s=0$, in agreement with
experiments\cite{experi}.  While the quantization of Hall drag is
expected to be violated for $T>0$ by activated processes contributing
to $\rho_{xx}^c$, the theory predicts that the spin Hall
resistivity, 
should vanish even at non-zero temperature for $T<T_{KT}$, which would
be interesting to explore experimentally.  For $d>d_c$, in phase III, we
obtain $C^{1,-1}=C^{-1,1}=0$ numerically, signaling the decoupling of
two layers.  As a consequence, the drag conductance and resistance
reduce to zero at $T=0$, which also agrees with the
experiment\cite{experi} at larger $d$ ($d>d_c^{exp})$.  Lastly, for the
relatively pure sample used in Ref.\onlinecite{experi}, phase I is
directly neighboring phase III which results in $\rho_{xx}^d\neq 0$ only
along the phase transition line $d=d_c$ at very low temperature, a
property again observed experimentally\cite{experi}.  For more
disordered samples at intermediate $d$, phase II (gauge glass)
intervenes.  While we expect this phase exhibits the same transport
coefficients as phase I at $T=0$, it has no associated
Kosterlitz-Thouless transition.  This implies that $\rho^s_{xy},
\rho^s_{xx}$ are generally non-zero for $T>0$ in this range, and
probably that $\rho_{xx}^d$ is enhanced at low but non-zero
temperatures.  We leave a more detailed investigation of the gauge glass
to a future study.

We would like to acknowledge helpful discussions with 
Jim Eisenstein, Matthew Fisher, Dung-Hai Lee,
and Z. Y. Weng.  This work was supported by  
ACS-PRF \# 36965-AC5 and Research Corporation Award 
CC5643, the NSF through grants DMR-00116566 (DNS),
DMR-9985255, the Sloan and Packard foundations (LB),
and DOE grant DE-FG02-99ER45747 (ZW).
DNS would  like to thank the hospitality and support from the KITP 
at Santa Barbara.

\end{document}